\documentclass[prb,twocolumn,showpacs,preprintnumbers,amsmath,amssymb]{revtex4}
\usepackage{dcolumn}
\usepackage{bm}
\usepackage{amsmath,graphicx}
\usepackage{color}
\usepackage{times}

\def\d{{\rm d}}
\def\dx{{\rm d}x}
\def\dy{{\rm d}y}
\def\e{{\rm e}}
\def\im{{\rm i}}
\def\vf{v_\mathrm{F}}
\def\ef{E_\mathrm{F}}

\begin{document}
\title{Density of states
anomalies in
multichannel quantum wires}

\author{Hideo Yoshioka}
\affiliation{Department of Physics, Nara Women's University, Nara 630-8506, Japan}

\author{Hiroyuki Shima}
\affiliation{
Division of Applied Physics, Faculty of Engineering,
Hokkaido University, Sapporo, Hokkaido 060-8628, Japan}

\date{\today}

%
%

\begin{abstract}
We reformulate the Tomonaga--Luttinger liquid theory
for quasi-one-dimensional Fermion systems with  many subbands across the Fermi energy.
Our theory enables  us to obtain a  rigorous expression of the local density of states (LDOS)
for general multichannel quantum wires,
describing how the power-law anomalies of LDOS depend on
inter- and intra-subbands couplings as well as the Fermi velocity of  each band.
The resulting formula for the exponents is valid
in the case of both bulk contact and edge contact,
and thus plays a fundamental role in the physical properties
of multicomponent Tomonaga--Luttinger liquid systems.
\end{abstract}





\maketitle

\section{Introduction}

Bosonization is one of the most powerful techniques
for describing 
the properties of
one-dimensional (1D) interacting electron systems.
In 1D systems, even a slight interaction between electrons strongly affects
the quantum nature,
resulting in the occurrence of  Tomonaga--Luttinger liquid (TLL) states.\cite{Tomonaga,Luttinger,Mattis_Lieb_1965,Haldane_1981}
TLL states exhibit power-law anomalies in physical quantities,
as predicted by the bosonization theory.\cite{Voit_1995}
A prominent example is the power-law singularity of
the single-particle density of states $D(E, T)$
near the Fermi energy, $\ef$,  represented by
$D(E, 0) \propto |E - \ef|^{\lambda}$ and
$D(\ef, T) \propto T^{\lambda}$, with
$E$ and $T$ being the energy and the temperature, respectively.
The value of $\lambda$, called the TLL exponent, is dependent on
the interaction strength\cite{Voit_1995}
and other parameters characterizing
the 1D system.\cite{Fabrizio_1995,Eggert_1996,Mattsson_1997,Bockrath_1999,Yao_1999,Postma_2000,Shima_2009,Bubanja_2009}
Recently, it has been suggested that a continuous variation in $\lambda$ can be  produced by an external field;\cite{Bellucci_2006,Klanjsek_2008,Shima_arXiv}
this implies artificial control of the transport properties of quasi-1D conductors,
since $\lambda$ governs the power-law behaviors of the differential tunneling conductance\cite{Bockrath_1999}
$dI/dV \propto |V|^{\lambda}$ at high bias voltages $(eV \gg k_B T)$
and the temperature-dependent conductance
$G(T) \propto T^{\lambda}$ at low voltages $(eV \ll k_B T)$.

Experimental realizations of TLL states encompass
various systems showing highly anisotropic conductivity:
metallic,\cite{Slot_2004,Venkataraman_2006}
semiconducting,\cite{Auslaender_2002,Auslaender_2005,Tserkovnyak_2002,Tserkovnyak_2003,Levy_2006,Steinberg_2008,Jompol_2009}
and organic nanowires,\cite{Schwartz_1998,Claessen_2002,Sing_2003,Aleshin_2004,Aleshin_2006}
and carbon nanotubes\cite{Bockrath_1999,Yao_1999,Bachtold_2001,Ishii_2003,Yoshioka_2003,Tombros_2006} are a few examples.
These actual quasi-1D conductors possess a finite cross-section,
thus exhibiting a finite number of transmission channels in the transverse direction
(except for a limited case in which $\ef$ is small enough for
only the lowest subband to be involved).
The presence of multiple channels at $\ef$ causes inter-subband scatterings.
Furthermore, different channels can have different Fermi velocities, i.e.,
the slope of the dispersion curve at $\ef$ (see Fig.~\ref{fig:band}),
and thus contributions from each channel to the TLL exponent
differ from  each other.
Theories of multichannel TLL have been developed for the Hubbard model in the presence
of an external magnetic field,\cite{Penc_1993, Kitazawa_2003,
Frahm_2008},
where the discrepancies
in the Fermi velocity between up- and down-spins
are taken into account.
A similar issue was also discussed in the study of quasi-1D Bose
gases.\cite{Tokuno2008}
The effect of inter-subband scattering on the TLL
exponent has been investigated in connection with
the TLL behavior of multiwall carbon nanotubes.\cite{Egger1999}
However, to the best of our knowledge, singular behavior in the density of states $D(E,T)$
remains unresolved for multichannel TLL systems with the coexistence of
inter-subband scatterings and Fermi-velocity variations.
Hence, the rigorous expression of $D(E,T)$ in multichannel TLL systems
is desirable for describing the
transport properties and photo-emission spectra
those will be experimentally observed in actual quasi-1D conductors.

In this paper, we reformulate the multichannel TLL theory
in order to derive the anomalous
energy- and temperature-dependences
of the local density of states of quasi-1D Fermion systems.
Cases of locations both far from the boundary and close to it, which correspond
to bulk contact and  end contact of  the transport properties, respectively, are discussed.
We demonstrate clearly how the TLL exponents of multichannel systems
depend on mutual interaction and Fermi velocities.
The resulting formula for the exponents, as well as the theoretical framework
we have established, will provide clues to exploiting the effects of subband couplings
and  Fermi velocity variations on the nature of TLLs in real 1D systems.

The paper is organized as follows.
In Section II, the multichannel TLL theory is developed for
$N$-channel quasi-1D Fermionic systems with different Fermi velocities.
The local densities of states far from and close to the boundary are
calculated in Section III.
As a simple example, in Section IV,
the case of spinless Fermions is discussed and
the $N$-dependences of the exponents are obtained for long-range
interaction limits.
The paper closes with a summary in Section V.
In the following, the unit  $\hbar = k_B = 1$ is used, unless explicitly
stated otherwise.

\section{Multicomponent Tomonaga--Luttinger liquids}

\subsection{Bosonization}

\begin{figure}[ttt]
\includegraphics[width=9.5cm]{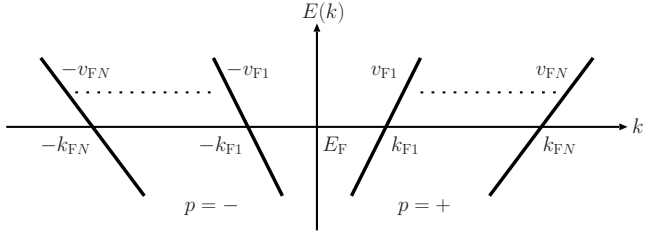}
\caption{Sketch of the energy dispersion of the present system, with
$N$ energy bands  cross the Fermi energy, $\ef$.
The Fermi velocity and the Fermi wavenumber of the $\nu$-th band ($\nu =
 1, \cdots, N$) are denoted by
$v_{{\rm F}\nu}$ and $k_{{\rm F}\nu}$, respectively.
The symbol $p=+$ ($-$) indicates a one-particle state moving
toward the right (left).}
\label{fig:band}
\end{figure}

We consider a quasi-1D Fermion system with
$N$ 1D energy bands cross the Fermi energy, $\ef$.
The band structure close to $\ef$
is schematically shown in Fig. \ref{fig:band}.
Here, the Fermi velocity and the Fermi wavenumber of the $\nu$-th band
($\nu = 1, \cdots, N$) are
denoted by $v_{{\rm F}\nu}$ and $k_{{\rm F}\nu}$, respectively, and
the one-particle state moving to the right (left)  are indicated by
$p=+$ ($-$).
The kinetic energy of the Hamiltonian, $\mathcal{H}_\mathrm{k}$, is
expressed by
\begin{align}
 \mathcal{H}_\mathrm{k} &= \sum_{\nu=1}^N \sum_{p=\pm} \sum_k p
 v_{\mathrm{F}\nu} k  c^\dagger_{k,p,\nu} c_{k,p,\nu},
\label{eqn:Hk}
\end{align}
where the one-particle energy and  the wavenumber $k$
are measured from $\ef$ and $pk_{\mathrm{F}\nu}$, respectively.
In Eq.~(\ref{eqn:Hk}),  $c^\dagger_{k,p,\nu}$ denotes the creation operator of the
Fermion with wavenumber $k$, branch $p$, and band index $\nu$.

Let us introduce
the density operator of the $p$-branch of the $\nu$-th band,
defined as
\begin{align}
\rho_{p,\nu}(q) \equiv
\begin{cases}
\sum_k c^\dagger_{k+q,p,\nu} c_{k,p,\nu} & \cdots \hspace{1em} \text{$q\neq 0$}\\
N_{p,\nu} = \sum_k : c^\dagger_{k,p,\nu} c_{k,p,\nu} : & \cdots
 \hspace{1em} \text{$q = 0$}
\end{cases}
,
\end{align}
which satisfies the commutation relation
$[\rho_{p,\nu}(-q), \rho_{p',\nu'}(q')] = \delta_{pp'}
\delta_{\nu \nu'} \delta_{q q'} pqL/(2\pi)$.
In terms of $\rho_{p,\nu}(q)$,
$\mathcal{H}_\mathrm{k}$
is expressed by\cite{Mattis_Lieb_1965,Haldane_1981}
\begin{align}
\mathcal{H}_\mathrm{k} =& \sum_{\nu = 1}^N \frac{\pi v_{{\rm F}\nu}}{L} \sum_{p,q}
\rho_{p,\nu}(q) \rho_{p,\nu}(-q), 
\end{align}
where $L$ is the length of the system.
The most general form of the mutual interaction between Fermions
leading to the $N$-component TLL
is written as\cite{Penc_1993}
\begin{align}
\mathcal{H}_\mathrm{int}
&= \frac{1}{2L} \sum_{\nu,\nu'=1}^N \sum_{p,q}
\Big\{
\tilde{g}_2 (\nu,\nu')
\rho_{p,\nu}(q) \rho_{-p,\nu'}(-q) \nonumber \\
&+ \tilde{g}_4 (\nu,\nu') \rho_{p,\nu}(q) \rho_{p,\nu'}(-q) \Big\}.
\label{eqn:A_int}
\end{align}
The matrix elements $\tilde{g}_2 (\nu,\nu')$ and $\tilde{g}_4
(\nu,\nu')$ depend on the details of the model we consider.
Specifically, the case with $\tilde{g}_2 \equiv \tilde{g}_4$
corresponds to the model for multiwall carbon nanotubes considered
in Ref.~\onlinecite{Egger1999}.
As an example, we will discuss the case of the spinless Fermion in
Section~\ref{sec:SF}.

We introduce the phase variables $\theta_\nu (x)$
and $\phi_\nu (x)$ $(\nu = 1, \cdots, N)$, defined as
\begin{align}
 \theta_\nu (x) =& - \frac{1}{\sqrt{2}} \sum_p p
\Bigg\{
 Q_{p,\nu} - \frac{2 \pi p x}{L} N_{p,\nu} \nonumber \\
&- \frac{2 \pi \im}{L} \sum_{q
 \neq 0} p 
\frac{\e^{-\im q x}}{q} \rho_{p,\nu}(q) - p \frac{\pi}{2} N_{-p,\nu}
\Bigg\}, \\
 \phi_\nu (x) =& - \frac{1}{\sqrt{2}} \sum_p
\Bigg\{
 Q_{p,\nu} - \frac{2 \pi p x}{L} N_{p,\nu} \nonumber \\
&- \frac{2 \pi \im}{L} \sum_{q
 \neq 0} p 
\frac{\e^{-\im q x}}{q} \rho_{p,\nu}(q) - p \frac{\pi}{2} N_{-p,\nu}
\Bigg\},
\end{align}
where $[Q_{p,\nu}, N_{p',\nu'}] = \im \delta_{pp'} \delta_{\nu \nu'}$.
In the summation in terms of $q$, the ultraviolet cutoff $\exp ( - \alpha
|q|/2)$ is implicitly included.
The phase variables satisfy the commutation relation,
$[ \theta_\nu(x), \phi_{\nu'} (x') ] = \im 2 \pi \delta_{\nu\nu'} \theta (x-x')$ for $L \to
\infty$, with $\theta (x)$ being the conventional step function.
In terms of the above phase variables,
the Hamiltonian is written as
\begin{align}
\mathcal{H} = \frac{1}{2} \sum_{\nu,\nu'=1}^N
\int \d x \left\{
\Pi_\nu ( K^{-1} )_{\nu\nu'} \Pi_{\nu'} + \partial_x \theta_\nu
 V_{\nu\nu'} \partial_x \theta_{\nu'}
\right\},
\label{eqn:phase_Hamiltonian}
\end{align}
where $\Pi_\nu = - \partial_x \phi_\nu/(2\pi)$.
This is the general form of the phase Hamiltonian expressing the $N$-component
TLL.
The symmetric matrices $K$ and $V$ are defined as follows:
\begin{align}
(K^{-1})_{\nu\nu'} &= 2 \pi v_{F \nu} \delta_{\nu\nu'} + \tilde{g}_4 (\nu,\nu')
 - \tilde{g}_2 (\nu,\nu'),  \label{eq_shima20} \\
V_{\nu\nu'} &= \frac{v_{F\nu}}{2\pi} \delta_{\nu\nu'} + \frac{\tilde{g}_4 (\nu,\nu')
 + \tilde{g}_2 (\nu,\nu') }{4\pi^2}. \label{eq_shima21}
\end{align}
The Fermion operator, defined by
$\psi_{p,\nu} (x) = 1/\sqrt{L} \sum_k \e^{\im k x} c_{k,p,\nu}$,
is related to the phase variables as
\begin{align}
 \psi_{p,\nu} (x) &= \frac{\eta_{\nu}}{\sqrt{2\pi\alpha}}
\exp\left\{ \im \frac{p}{\sqrt{2}} \left[ \theta_\nu(x) + p \phi_\nu
 (x)\right] \right\},
\label{eqn:field}
\end{align}
where $\eta_{\nu}$ expresses the Majorana Fermion operator satisfying
$\eta_{\nu} = \eta_{\nu}^\dagger$ and
$\left\{ \eta_{\nu}, \eta_{\nu'}\right\} = 2
\delta_{\nu \nu'}$.

\subsection{Diagonalization}
The Hamiltonian given by Eq.~(\ref{eqn:phase_Hamiltonian}) has a bilinear form
with respect to $\partial_x \phi_{\nu}$ and $\partial_x \theta_{\nu}$,
and thus can be diagonalized by the standard unitary transformation,
as shown below.

The equations of motion of the phase variables derived from
Eq.~(\ref{eqn:phase_Hamiltonian}) read as
\begin{align}
 \frac{\partial}{\partial t} \bm{\Pi} = V \frac{\partial^2}{\partial
 x^2} \bm{\theta}, \\
 \frac{\partial}{\partial t} \bm{\theta} = K^{-1} \bm{\Pi},
\end{align}
with $\bm{\theta} = (\theta_1,\theta_2,\cdots,\theta_N)^\mathrm{T}$ and
$\bm{\Pi} = (\Pi_1,\Pi_2,\cdots,\Pi_N)^\mathrm{T}$.
Here, the energy eigenvalue, $\omega = v |k|$,
and the eigenvector $\bm{X}$ corresponding to it are determined by
\begin{align}
 (v^2 K - V) \bm{X} = 0, \label{eq_shima25}
\end{align}
whose solutions are denoted by $v_j$ and $\bm{X}_j$ ($j=1,2,\cdots,N$).
The eigenvectors are normalized as
$(\bm{X}_i, K \bm{X}_j) = \delta_{ij}$.

To obtain a concise representation of $\mathcal{H}$,
we define the unitary transformation as
\begin{align}
 \bm{\theta} &= X \bm{\Theta}, \\
 \bm{\Pi} &= K X \bm{\Xi},
\end{align}
where
$\bm{\Theta} = (\Theta_1, \Theta_2, \cdots, \Theta_N)^\mathrm{T}$ and
$\bm{\Xi} = (\Xi_1, \Xi_2, \cdots, \Xi_N)^\mathrm{T}$.
The $N \times N$ matrix $X$ consists of the set of eigenvectors $\bm{X}_j$
as $X = (\bm{X}_1, \bm{X}_2, \cdots, \bm{X}_N)$, and
satisfies $X^\mathrm{T} KX = 1$.
Under the transformation,
$[\Theta_j(x), \Xi_{j'}(x')] = \im \delta_{jj'} \delta (x-x')$.
By using the new variables, we obtain an alternative form of $\mathcal{H}$,
given by
\begin{align}
 \mathcal{H} &= \frac{1}{2} \sum_{j=1}^N \int \d x
\left\{ \Xi_j^2 + v_j^2 (\partial_x \Theta_j)^2 \right\},
\label{eqn:a4}
\end{align}
and that of the field operator defined in Eq.~(\ref{eqn:field}),
\begin{align}
 \psi_{p,\nu}(x) &= \frac{\eta_{\nu}}{\sqrt{2 \pi \alpha}} \nonumber \\
&\times \exp \left( \im \frac{p}{\sqrt{2}} \sum_{j=1}^N
\left\{
X_{\nu j} \Theta_j (x) + p (KX)_{\nu j} \Phi_j (x)
\right\}
\right),
\end{align}
where $\phi_\nu = \sum_{j=1}^N (KX)_{\nu j} \Phi_j$.

\section{Density of States}

In this section, we discuss the local density of states $D(\omega,T,x)$ with $\omega \equiv E-\ef$
for $\omega\ll \ef$, where $x$ denotes the position along the 1D direction.
As noted earlier, the TLL exponent that characterizes the singularity
of $D(\omega,T,x)$ near $\ef$ is $x$ dependent.
From a practical view,
it is specifically interesting to study the semi-infinite system
with its end at the origin\cite{Fabrizio_1995,Eggert_1996,Mattsson_1997}
and discuss the cases with $x \to 0$ and $x \to \infty$, which
correspond to the end contact and the bulk contact, respectively.
In the following argument, we therefore  derive the TLL exponent for both cases,
as well as the explicit forms of $D(\omega,T,x)$ as functions of $\omega$ and $T$.

The local density of states is given by the summation of the contribution from  each
band: $D(\omega,T,x) =
\sum_{\nu = 1}^N D_{\nu} (\omega,T,x)$.
The contribution from the $\nu$-th band,
$D_{\nu}(\omega,T,x)$, is given by
\begin{align}
 D_{\nu} (\omega, T, x) &= \frac{1}{2\pi} \int_{-\infty}^{\infty} \d t
\e^{\im \omega t}
\langle \left\{
\psi^\dagger_{\nu}(x,0), \psi_{\nu} (x,t)
\right\} \rangle,
\label{eqn:Dpnu}
\end{align}
where $\psi_{\nu} (x,t) = \e^{\im k_{\mathrm{F}\nu}x} \psi_{+,\nu} (x,t)
+ \e^{- \im k_{\mathrm{F}\nu}x} \psi_{-,\nu} (x,t)$.
Here, $\left\{A,B\right\} \equiv AB + BA$ and
$\langle \cdots \rangle$ means the thermal average.
The quantity in the integrand in Eq.~(\ref{eqn:Dpnu}) for $x \to \infty$
is proven to be
\begin{align}
& \langle \left\{
\psi^\dagger_{\nu}(x \to \infty,0), \psi_{\nu} (x \to \infty,t)
\right\} \rangle \nonumber \\
=& \frac{1}{\pi \alpha}\left[F^{(b)}(t) + F^{(b)}(-t) \right],
 \label{eqn:a1}\\
F^{(b)}(t) 
=& \left( \frac{\pi T t}{\sinh \pi T t}
 \right)^{\sum_{j=1}^N Y_{\nu,j}^{(b)} } \times \prod_{j=1}^N
\frac{1}{\left(1 - \im v_j t / \alpha \right)^{Y_{\nu,j}^{(b)} }}, \label{eqn:a2}\\
Y_{\nu,j}^{(b)} =& \frac{1}{2} \left\{ \frac{(X_{\nu,j})^2}{2 \pi v_j} +
 2 \pi v_j \left[(KX)_{\nu,j}
 \right]^2\right\}, \label{eqn:a3}
\end{align}
where the superscript $(b)$ means the case of a ``bulk" contact.
Similarly, the counterpart for the ``edge" contact case  $x\to 0$,
labeled by $(e)$,
obeys Eqs.~(\ref{eqn:a1})--(\ref{eqn:a3}) with $Y_{\nu,j}^{(b)}$ replaced
by
\begin{align}
Y_{\nu,j}^{(e)} =&
 2 \pi v_j \left[(KX)_{\nu,j}\right]^2.
\label{eqn:aa3dash}
\end{align}
The derivations of Eqs.~(\ref{eqn:a1})--(\ref{eqn:aa3dash})
are shown in Appendix A.

We are ready to obtain the rigorous expression for
$D_\nu^{(b/e)} (\omega,T) \equiv D_\nu(\omega,T,x \to \infty/0)$.\cite{Mattsson_1997}
Since $D^{(b/e)}_{\nu} (0,0)$ vanishes as long as $\sum_{j=1}^N Y^{(b/e)}_{\nu,j} > 1$,
we subtract it from the result and take the limit $\alpha \to 0$.
Eventually, we attain the desired formulas:
\begin{align}
 & D_{\nu}^{(b/e)} (\omega,T) \nonumber \\
=& \frac{1}{2 \pi^2 \alpha} \int_{-\infty}^\infty \d t
\left\{
\e^{\im \omega t} \left( \frac{\pi Tt}{\sinh \pi Tt} \right)^{\sum_{j=1}^N
 Y_{\nu,j}^{(b/e)}} -1
\right\} \nonumber \\
&\times \left\{
\prod_{j=1}^N \frac{1}{\left(1 - \im v_j t / \alpha \right)^{Y_{\nu,j}^{(b/e)}}}
+ \prod_{j=1}^N \frac{1}{\left(1 + \im v_j t / \alpha \right)^{Y_{\nu,j}^{(b/e)}}}
\right\} \nonumber \\
=& \frac{2}{\pi^2 \alpha} \prod_{j=1}^N
 \left(\frac{\alpha}{v_j}\right)^{Y_{\nu,j}^{(b/e)}} \cos \left(\frac{\pi}{2}
 \sum_{j=1}^N Y_{\nu,j}^{(b/e)}\right) \nonumber \\
&\times
\int_0^\infty \d t
\left\{
\cos \omega t \left( \frac{\pi Tt}{\sinh \pi Tt} \right)^{\sum_{j=1}^N
 Y_{\nu,j}^{(b/e)}} -1
\right\}
\frac{1}{t^{\sum_{j=1}^N Y_{\nu,j}^{(b/e)}}}. \label{eq_23temp}
\end{align}
Equation (\ref{eq_23temp}) implies that
the $\omega$- and $T$-dependences of $D_{\nu}^{(b/e)} (\omega,T)$ are given by
\begin{align}
 D_{\nu}^{(b/e)} (\omega,0) =& \frac{1}{\pi \alpha}
\prod_{j=1}^N \left( \frac{\alpha}{v_j}\right)^{Y_{\nu,j}^{(b/e)}}
 \nonumber \\
&\times \frac{1}{\Gamma \left[ \sum_{j=1}^N Y_{\nu,j}^{(b/e)}\right]} \omega^{\sum_{j=1}^N
 Y_{\nu,j}^{(b/e)}-1}, \\
D_{\nu}^{(b/e)} (0,T) =& \frac{1}{\pi^2 \alpha}
\prod_{j=1}^N \left( \frac{\alpha}{v_j}\right)^{Y_{\nu,j}^{(b/e)}}
 \nonumber \\
&\times  \frac{ \left\{\Gamma \left[\sum_{j=1}^N Y_{\nu,j}^{(b/e)}/2 \right] \right\}^2}
{\Gamma\left[\sum_{j=1}^N Y_{\nu,j}^{(b/e)}\right]}
(2 \pi T)^{\sum_{j=1}^N Y_{\nu,j}^{(b/e)}-1},
\end{align}
where $\Gamma[z]$ is the Gamma function.
It thus follows that the TLL exponent
associated with the $\nu$-th band reads as
\begin{align}
 \lambda^{(b)}(\nu) &= \sum_{j=1}^N Y_{\nu,j}^{(b)} -1 \nonumber \\
&= \sum_{j=1}^N \frac{1}{2} \left\{ \frac{(X_{\nu,j})^2}{2 \pi v_j} + 2 \pi v_j
 [(KX)_{\nu,j}]^2 \right\} - 1,
\label{eqn:DOSpower-b}
\end{align}
for the bulk position, and
\begin{align}
 \lambda^{(e)}(\nu) &= \sum_{j=1}^N Y_{\nu,j}^{(e)} -1 \nonumber \\
&= \sum_{j=1}^N 2 \pi v_j [(KX)_{\nu,j}]^2 - 1,
\label{eqn:DOSpower-e}
\end{align}
for the edge.
Equations (\ref{eqn:DOSpower-b}) and (\ref{eqn:DOSpower-e})
are the main findings of this article.

Since $D(\omega,T,x)$ is given by the summation of the contributions
from each of the bands whose powers differ,
the smallest values would be observed in actual experiments,
such as photo-emissions and transport properties.

Specifically, if
$N=2$
with $v_{F1} = v_{F2} = v_F$,
the present model is reduced to the conventional
1D electron system, 
where the backward scattering between the different spins and the Umklapp
scattering are ignored.
In fact, by parameterizing as
\begin{align}
 \tilde{g}_2 (1,1) &= \tilde{g}_2 (2,2) = g_{2 \parallel} - g_{1 \parallel}, \\
 \tilde{g}_2 (1,2) &= \tilde{g}_2 (2,1) = g_{2 \perp}, \\
 \tilde{g}_4 (1,1) &= \tilde{g}_4 (2,2) = g_{4 \parallel}, \\
 \tilde{g}_4 (1,2) &= \tilde{g}_4 (2,1) = g_{4 \perp},
\end{align}
Eqs.~(\ref{eqn:DOSpower-b}) and
(\ref{eqn:DOSpower-e}) lead to the familiar forms
\begin{align}
 \lambda^{(b)}(1) = \lambda^{(b)}(2) &= \frac{1}{4} \left( K_\rho + K_\rho^{-1} +
 K_\sigma + K_\sigma^{-1}\right) -1, \\
 \lambda^{(e)}(1) = \lambda^{(e)}(2) &= \frac{1}{2} \left( K_\rho^{-1} +
  K_\sigma^{-1}\right) -1,
\end{align}
with
\begin{align}
 K_\rho &= \sqrt{
\frac
{2\pi\vf+g_{4\parallel}+g_{4\perp}-g_{2\parallel}-g_{2\perp}+g_{1\parallel}}
{2\pi\vf+g_{4\parallel}+g_{4\perp}+g_{2\parallel}+g_{2\perp}-g_{1\parallel}}
}, \\
 K_\sigma &= \sqrt{
\frac
{2\pi\vf+g_{4\parallel}-g_{4\perp}-g_{2\parallel}+g_{2\perp}+g_{1\parallel}}
{2\pi\vf+g_{4\parallel}-g_{4\perp}+g_{2\parallel}-g_{2\perp}-g_{1\parallel}}
}.
\end{align}

\section{$N$-channel spinless Fermions} \label{sec:SF}

In this section,
we derive the matrix elements of the mutual
interactions, which are included in the matrices $K$ and $V$ in Eqs.~
(\ref{eq_shima20}) and (\ref{eq_shima21}).
As a simple example,
we consider a quasi-1D spinless Fermion system where
$N$ 1D energy bands cross the Fermi energy, $\ef$.
In addition, in order to clarify the effects of the number of channels
on the exponents, those for the case of  long-range interaction are
derived.

The mutual interaction $\mathcal{H}_\mathrm{int}$ of the spinless
Fermion  can be expressed generally as
\begin{align}
 {\cal H}_\mathrm{int} = \frac{1}{2} \iint \d x \d x'
\psi^\dagger (x) \psi^\dagger (x') V(|x-x'|) \psi (x') \psi (x),
\label{eqn:Hint}
\end{align}
with $\psi (x)$ being the annihilation operator of the spinless Fermion.
Since we are discussing low-energy physics,
the interaction processes among the particles close
to $\ef$ are necessary.
In order to obtain such interaction processes,
the operator $\psi(x)$ is expanded, using the eigenfunctions of the
states across $\ef$, $\phi_{\nu,K}(x)$, as
\begin{align}
 \psi (x) &= \sum_{\nu = 1}^N \sum_{K} a_{\nu,K} \phi_{\nu,K}(x),
\label{eqn:expand}
\end{align}
where $a_{\nu,K}$ is the operator of the spinless Fermion
with the eigenstate $(\nu,K)$.
By inserting Eq.~(\ref{eqn:expand}) into Eq.~(\ref{eqn:Hint}),
$\mathcal{H}_\mathrm{int}$ is
expressed as
\begin{align}
\mathcal{H}_\mathrm{int} =& \frac{1}{2}
\sum_{\nu_1,\nu_2,\nu_3,\nu_4} \sum_{K_1, K_2, K_3, K_4}
V_{\nu_1K_1,\nu_2K_2;\nu_3K_3,\nu_4K_4} \nonumber \\
&\times a^\dagger_{\nu_1,K_1} a^\dagger_{\nu_2,K_2} a_{\nu_3,K_3} a_{\nu_4,K_4},
\label{eqn:Hint-expand}
\end{align}
where the matrix element of the mutual interaction is written as
\begin{align}
V_{\nu_1K_1,\nu_2K_2;\nu_3K_3,\nu_4K_4}
=& \iint \dx \dy V(|x-y|) \nonumber \\
& \hspace*{-3em} \times
\phi^*_{\nu_1,K_1} (x) \phi^*_{\nu_2,K_2} (y)
\phi_{\nu_3,K_3} (y) \phi_{\nu_4,K_4} (x).
\label{eqn:element_general}
\end{align}
We note that
as a result of momentum conservation,
the relation
$K_1 + K_2 - K_3 - K_4 = n G$ holds, where
$G$ is the reciprocal lattice vector and $n$ is an integer.
In the following, we discuss the case where the filling of each band
is incommensurate.
Then, only the normal processes satisfying $n=0$ are taken into
account as
\begin{align}
 V_{\nu_1K_1,\nu_2K_2;\nu_3K_3,\nu_4K_4}
 &= \delta_{K_1 + K_2, K_3+K_4} \nonumber \\
 & \times V_{\nu_1,\nu_2;\nu_3,\nu_4} (K_1,K_2;K_3,K_4).
\end{align}
In this case, $\mathcal{H}_\mathrm{int}$ is expressed by
\begin{align}
\mathcal{H}_\mathrm{int} =& \frac{1}{2}
\sum_{\nu_1,\nu_2,\nu_3,\nu_4} \sum_{p_1, p_2, p_3, p_4}
\sum_{k_1, k_2, k_3, k_4} \nonumber \\
&\times \delta_{p_1 k_{\mathrm{F}\nu_1}+p_2 k_{\mathrm{F}\nu_2},p_3 k_{\mathrm{F}\nu_3}+p_4 k_{\mathrm{F}\nu_4}}
\delta_{k_1+k_2,k_3+k_4} \nonumber \\
&\times
 V_{\nu_1,\nu_2;\nu_3,\nu_4}
 (p_1k_{\mathrm{F}\nu_1},p_2k_{\mathrm{F}\nu_2};p_3k_{\mathrm{F}\nu_3},p_4k_{\mathrm{F}\nu_4}) \nonumber \\
&\times c^\dagger_{k_1,p_1,\nu_1} c^\dagger_{k_2,p_2,\nu_2}
 c_{k_3,p_3,\nu_3}  c_{k_4,p_4,\nu_4},
\label{eqn:Hint-Bloch-N}
\end{align}
where $c_{p,k,\nu} = a_{\nu,p k_{\mathrm{F}_\nu}+k}$.
Here, the relations $K_i = p_i k_{\mathrm{F}\nu_i} + k_i$ and $|k_i| \ll
k_{\mathrm{F}\nu_j}$ $(i,j=1,\cdots,N)$ are used.
Assuming $k_{\mathrm{F}\nu} \neq k_{\mathrm{F}\nu'}$
for $\nu \neq \nu'$, Eq.~(\ref{eqn:Hint-Bloch-N}) is written as
$\mathcal{H}_\mathrm{int} = \mathcal{H}_\mathrm{int,1} +
\mathcal{H}_\mathrm{int,2} + \mathcal{H}_\mathrm{int,4}$,
where
\begin{align}
\mathcal{H}_\mathrm{int,1} =& \frac{1}{2}
\sum_{k,k',q} \sum_{p=\pm} \sum_{\nu,\nu'=1}^N
\!\!\! V_{\nu,\nu';\nu,\nu'} (p k_{\mathrm{F}\nu}, -p k_{\mathrm{F}\nu'};p
 k_{\mathrm{F}\nu }, -p k_{\mathrm{F}\nu'}) \nonumber \\
&\times
c^\dagger_{k+q,p,\nu} c^\dagger_{k'-q,-p,\nu'} c_{k',p,\nu}
 c_{k,-p,\nu'},
\label{eqn:H1}
\\
\mathcal{H}_\mathrm{int,2} =& \frac{1}{2}
\sum_{k,k',q} \sum_{p=\pm} \sum_{\nu,\nu'=1}^N
\!\!\! V_{\nu,\nu';\nu',\nu} (p k_{\mathrm{F}\nu}, -p k_{\mathrm{F}\nu'};-p
k_{\mathrm{F}\nu'}, p k_{\mathrm{F}\nu}) \nonumber \\
&\times
c^\dagger_{k+q,p,\nu} c^\dagger_{k'-q,-p,\nu'} c_{k',-p,\nu'}
 c_{k,p,\nu}, \label{eqn:H2}
\\
\mathcal{H}_\mathrm{int,4} =& \frac{1}{2}
\sum_{k,k',q} \sum_{p=\pm} \sum_{\nu=1}^N
V_{\nu,\nu;\nu,\nu} (p k_{\mathrm{F}\nu}, p k_{\mathrm{F}\nu};p
 k_{\mathrm{F}\nu}, p k_{\mathrm{F}\nu}) \nonumber \\
&\times c^\dagger_{k+q,p,\nu} c^\dagger_{k'-q,p,\nu} c_{k',p,\nu}
 c_{k,p,\nu}
\nonumber \\
+& \frac{1}{2}
\sum_{k,k',q} \sum_{p=\pm} \sum_{\nu \neq \nu'}^N
\Big\{
V_{\nu,\nu';\nu',\nu} (p k_{\mathrm{F}\nu}, p k_{\mathrm{F}\nu'};p
 k_{\mathrm{F}\nu'}, p k_{\mathrm{F}\nu}) \nonumber \\
&\times c^\dagger_{k+q,p,\nu} c^\dagger_{k'-q,p,\nu'} c_{k',p,\nu'}
 c_{k,p,\nu}
\nonumber \\
&+  V_{\nu,\nu';\nu,\nu'} (p k_{\mathrm{F}\nu}, p k_{\mathrm{F}\nu'};p
 k_{\mathrm{F}\nu}, p k_{\mathrm{F}\nu'}) \nonumber \\
&\times c^\dagger_{k+q,p,\nu} c^\dagger_{k'-q,p,\nu'} c_{k',p,\nu}
 c_{k,p,\nu'}
\Big\}.
\label{eqn:H4}
\end{align}
%
Here,
$\mathcal{H}_\mathrm{int,1}$
represents the backward scattering,
$\mathcal{H}_\mathrm{int,2}$ denotes
the forward scattering among the different branches,
and
$\mathcal{H}_\mathrm{int,4}$ expresses
the forward scattering between the same branches.
It should be noted that we neglect accidental situations in the momentum conservation,
for example,
$k_{\mathrm{F}\nu_1} - k_{\mathrm{F}\nu_2} = - k_{\mathrm{F}\nu_3} + k_{\mathrm{F}\nu_4}$
with $k_{\mathrm{F}\nu_1} \neq k_{\mathrm{F}\nu_4}$ and
$k_{\mathrm{F}\nu_2} \neq k_{\mathrm{F}\nu_3}$, in
the forward scattering among different branches.
Equations (\ref{eqn:H1}), (\ref{eqn:H2}), and (\ref{eqn:H4})
are reduced to
\begin{align}
\mathcal{H}_\mathrm{int}
=& \frac{1}{2L} \sum_{k,k',q} \sum_{p=\pm} \sum_{\nu,\nu'=1}^N \nonumber \\
\times &\Big\{ g_1 (\nu,\nu') c^\dagger_{k+q,p,\nu} c^\dagger_{k'-q,-p,\nu'} c_{k',p,\nu'} c_{k,-p,\nu}
\nonumber \\
&+ g_2 (\nu,\nu')
c^\dagger_{k+q,p,\nu} c^\dagger_{k'-q,-p,\nu'} c_{k',-p,\nu'}
 c_{k,p,\nu} \nonumber \\
&+ g_4 (\nu,\nu')
c^\dagger_{k+q,p,\nu} c^\dagger_{k'-q,p,\nu'} c_{k',p,\nu'} c_{k,p,\nu}
\Big\}, 
\label{eqn:AA_int}
\end{align}
where
\begin{align}
 g_1 (\nu,\nu') =& L V_{\nu,\nu';\nu,\nu'}
(k_{\mathrm{F}\nu}, -k_{\mathrm{F}\nu'};k_{\mathrm{F}\nu }, - k_{\mathrm{F}\nu'}),
\label{eq_shima12}
\\
g_2 (\nu,\nu') =& L V_{\nu,\nu';\nu',\nu}
(k_{\mathrm{F}\nu}, -k_{\mathrm{F}\nu'};-k_{\mathrm{F}\nu'}, k_{\mathrm{F}\nu}), \label{eq_shima13} \\
g_4 (\nu,\nu') =& L \big\{
V_{\nu,\nu;\nu,\nu}
(k_{\mathrm{F}\nu}, k_{\mathrm{F}\nu};k_{\mathrm{F}\nu}, k_{\mathrm{F}\nu})  \delta_{\nu,\nu'} \nonumber \\
+& [ V_{\nu,\nu';\nu',\nu}
(k_{\mathrm{F}\nu}, k_{\mathrm{F}\nu'};k_{\mathrm{F}\nu'}, k_{\mathrm{F}\nu}) \nonumber \\
&-
V_{\nu,\nu';\nu,\nu'}
(k_{\mathrm{F}\nu}, k_{\mathrm{F}\nu'};k_{\mathrm{F}\nu}, k_{\mathrm{F}\nu'}) ](1-\delta_{\nu,\nu'})
\big\}.  \label{eq_shima14}
\end{align}
Here, the relation
$\phi_{\nu,K}^*(x) = \phi_{\nu, -K}(x)$, which is a result of time-reversal symmetry, is used.
Note that $g_1(\nu,\nu')$,  $g_2(\nu,\nu')$, and
$g_4(\nu,\nu')$ are the real symmetric matrices.
By comparing Eq.~(\ref{eqn:AA_int}) with Eq.~(\ref{eqn:A_int}),
we obtain
$\tilde{g}_2(\nu,\nu') = g_2 (\nu,\nu') - g_1 (\nu,\nu')$ and
$\tilde{g}_4(\nu,\nu') = g_4 (\nu,\nu')$.

Here, we consider the case where the Fermi velocities of all the
channels are equal to each other, i.e., $v_{\mathrm{F}\nu} =
v_\mathrm{F}$.
In addition, we assume that the matrix elements are given by  $g_2(\nu,\nu') =
g_4(\nu,\nu') = g$ and $g_1(\nu,\nu')=0$, which is a simple but effective
approximation for long-range interactions.\cite{Egger1999}
Here, the velocities of the excitation are obtained as $v_1 = v_\mathrm{F}
\sqrt{1+ Ng/(\pi \vf)}$ and $v_j = \vf$ ($j=2,\cdots,N$).
The eigenvector corresponding to $v=v_1$ is written as
$\bm{X}_1 = \sqrt{2 \pi v_F/N} (1,\cdots,1)^\mathrm{T}$.
As a result, the exponents are obtained as follows:
\begin{align}
\lambda^{(b)}(\nu) &= \frac{1}{N} \left\{ \frac{1}{2}
 \left(\frac{\vf}{v_1} + \frac{v_1}{\vf} \right) -1 \right\}, \\
\lambda^{(e)}(\nu) &= \frac{1}{N}
 \left(\frac{v_1}{\vf} -1 \right).
\end{align}
Thus, the TLL exponents for both locations decrease with increasing
$N$, and  are proportional to $N^{-1/2}$ for $N \gg 1$.

\section{Conclusion}
In the present paper, we reformulated the TLL theory
for multichannel 1D Fermion systems.
The theory obtained enables derivation of rigorous expressions
for the local density of states and the corresponding TLL exponents,
$\lambda^{(b/e)}(\nu)$, with respect to the $\nu$-th band.
The strategy for evaluating $\lambda^{(b/e)}(\nu)$
is summarized as follows:
\begin{enumerate}
\item
Define the functional forms of the 1D eigenfunction $\phi_{\nu,K}(x)$ and the interaction $V(|x-y|)$
appropriate for the system being considered.
\item
Calculate $V_{\nu_1K_1,\nu_2K_2;\nu_3K_3,\nu_4K_4}$ using Eq.~(\ref{eqn:element_general}).
\item
Using the above result, set the mutual interaction terms, $\tilde{g}_i(\nu,\nu')$,
that are necessity to define the interaction $\mathcal{H}_{\rm int}$ given by Eq.~(\ref{eqn:A_int}).
Particularly when considering spinless Fermions,
we can obtain $g_i(\nu,\nu')$ for $i=1,2,4$
by substituting the results of step 2 into Eqs.~(\ref{eq_shima12})--(\ref{eq_shima14}).
\item
Set $(K^{-1})_{\nu,\nu'}$ and $V_{\nu,\nu'}$ according to Eqs.~(\ref{eq_shima20}) and (\ref{eq_shima21}).
\item
Solve the eigenvalue problem (\ref{eq_shima25}) to obtain $v_j$ and $\bm{X}_j$ for $j=1,\cdots,N$.
\item
Evaluate $Y_{\nu,j}^{(b)}$ and $Y_{\nu,j}^{(e)}$ from
Eqs.~(\ref{eqn:a3}) and (\ref{eqn:aa3dash}).
\item
Finally, we obtain the exponents $\lambda^{(b)}(\nu)$ and
$\lambda^{(e)}(\nu)$ from Eqs.~(\ref{eqn:DOSpower-b}) and
(\ref{eqn:DOSpower-e}).
\end{enumerate}
By applying the strategy for $N$-channel spinless Fermions with long-range interactions,
we have revealed that both TLL exponents approach  zero
in proportion to $N^{-1/2}$ for $N \gg 1$.\cite{Egger1999}
This finding implies that the power-law anomalies
disappear and the energy-independent density of states, which is
a manifestation of the realization of Fermi liquids, emerges for $N \to \infty$.

Before closing, we remark that
the present theory began with the electronic Hamiltonian of the mutual interaction in
Eq.~(\ref{eqn:A_int}), which leads to the bosonic Hamiltonian
Eq.~(\ref{eqn:phase_Hamiltonian}) where no nonlinear terms exist.
Even in the presence of interaction terms leading to nonlinear
terms, the present theory can be useful if the the nonlinear
terms are renormalized to zero.
In such cases, it is necessary to take account of the renormalization of
the parameters $K$ and $V$ by the diminishing nonlinear terms.

\section*{Acknowledgment}
This work was supported by Nara Women's University Intramural Grant for
Project Research and a Grant-in-Aid for Scientific Research
(Nos. 22540329 and 22760058) from the Ministry of Education, Culture, Sports, Science
and Technology.
HS acknowledges financial support by The Sumitono Foundation.

\appendix
\section{Derivation of Eqs.~(\ref{eqn:a1})--(\ref{eqn:aa3dash})}
We discuss a semi-infinite system with its end at the origin.
For convenience, we scale the bosonic fields as
\begin{align}
 \tilde{\Theta}_j(x,t) &= \sqrt{v_j} \Theta_j(x,t), \\
 \tilde{\Phi}_j(x,t) &= \frac{1}{ 2 \pi \sqrt{v_j}} \Phi_j(x,t),
\end{align}
where $[\tilde{\Theta}_j(x,t), \tilde{\Phi}_{j'}(y,t)] = \im
\delta_{jj'} \theta(x-y)$.
By using field operators, the Hamiltonian is written as
\begin{align}
\mathcal{H} = \sum_{j=1}^N \frac{v_j}{2} \int \d x \left\{ \tilde{\Xi}_j^2 +
(\partial_x \tilde{\Theta}_j)^2 \right\},
\end{align}
where $\tilde{\Xi}_j = - \partial_x \tilde{\Phi}_j$.
The boundary condition at the origin requires the Fermion field for the
$\nu$-th subband $\psi_\nu (0) = 0$,
i.e., $\psi_{-,\nu}(0) = - \psi_{+,\nu}(0)$.
This condition leads to
\begin{align}
 \frac{1}{\sqrt{2}} \sum_{j=1}^N \frac{X_{\nu,j}}{\sqrt{v_j}}
 \tilde{\Theta}_j (0,t) = \left( n + \frac{1}{2}\right) \pi,
\end{align}
with $n$ being an arbitrary integer.

The mode expansion, together with the canonical quantization, leads to
\begin{align}
 \tilde{\Theta}_j(x,t) &= C_j + \tilde{\Theta}'_j(x,t), \label{eqn:tit}\\
 \tilde{\Theta}'_j(x,t) &= \frac{1}{\pi} \int_0^\infty \d q \frac{\sin
 qx}{\sqrt{q}} \left\{ - \im e^{-\im v_j q t} b_j (q) + \im e^{\im v_j q
 t} b_j^\dagger (q) \right\}, \label{eqn:titd} \\
 \tilde{\Phi}_j(x,t) &= - \frac{1}{\pi} \int_0^\infty \d q \frac{\cos
 qx}{\sqrt{q}} \left\{ e^{-\im v_j q t} b_j (q) + e^{\im v_j q
 t} b_j^\dagger (q) \right\}, \label{eqn:tip}
\end{align}
where $C_j$ is the c-number satisfying
$(1/\sqrt{2}) \sum_{j=1}^N (X_{\nu,j}/\sqrt{v_j})
 C_j = \left\{ n + ({1}/{2}) \right\} \pi$, and $b_j(q)$ is the bosonic
 operator with
$[b_j(q),b_{j'}^\dagger(q')] = \pi \delta_{jj'} \delta (q-q')$.
The ultraviolet cutoff $\exp (- \alpha q/2)$ is inserted if necessary
in the $q$-integral in Eqs.~(\ref{eqn:titd}) and (\ref{eqn:tip}).
Note that $\partial_t \tilde{\Theta}_j (x,t) = - v_j \partial_x
\tilde{\Phi}_j (x,t)$.
As a result of  Eqs.~(\ref{eqn:tit})--(\ref{eqn:tip}), the Fermion field of the $\nu$-th band satisfies $\psi_{-,\nu} (x,t) = -
\psi_{+,\nu} (-x,t)$.
The Hamiltonian is written as
\begin{align}
\cal{H} &= \sum_{j=1}^N \frac{1}{\pi} \int_0^\infty \d q v_j q
 b_j^\dagger(q) b_j(q).
\label{eqn:HHH}
\end{align}

The quantity $\langle \left\{ \psi^\dagger_{\nu} (x,0), \psi_{\nu} (x,t) \right\}\rangle$
is calculated as follows:
\begin{align}
 & \langle \left\{ \psi^\dagger_{\nu} (x,0), \psi_{\nu} (x,t)
 \right\}\rangle \nonumber \\
\simeq& \langle \left\{ \psi_{+,\nu}^\dagger (x,0), \psi_{+,\nu}(x,t)
 \right\} \rangle
+ (x \to -x) \nonumber \\
=& \frac{1}{2 \pi \alpha} \left\{ \prod_{j=1}^N G_{\nu,j} (x,t)  H_{\nu,j}(x,t) +
 \prod_{j=1}^N G_{\nu,j} (x,t) H^{-1}_{\nu,j}(x,t) \right\} \nonumber \\
&+ (x \to -x),
\end{align}
where
\begin{align}
 G_{\nu,j} (x,t) =&
\left\langle
\exp
\left\{
- \im \frac{1}{\sqrt{2}}
\left( f_{\nu,j}(x,0) - f_{\nu,j}(x,t) \right)
\right\}
\right\rangle \nonumber \\
=& \exp \left\{
- \frac{1}{4} \left\langle \left( f_{\nu,j}(x,0) - f_{\nu,j}(x,t) \right)^2 \right\rangle
\right\},  \\
H_{\nu,j} (x,t) =&
\exp \left\{
\frac{1}{4}
\left[
f_{\nu,j}(x,0), f_{\nu,j}(x,t)
\right]
\right\},
\end{align}
with
$f_{\nu,j}(x,t)  = X_{\nu,j}/\sqrt{v_j} \tilde{\Theta}'_j (x,t) + 2 \pi \sqrt{v_j} (KX)_{\nu,j} \tilde{\Phi}_j
(x,t)$.
Here, we ignore the rapidly oscillating terms proportional to
$\exp(\pm \im 2 k_{\mathrm{F}\nu} x)$ because these contributions can probably not be
observed directly due to averaging over several lattice sites in the
experiments.
From Eqs.~(\ref{eqn:titd}) and (\ref{eqn:tip}), together with (\ref{eqn:HHH}),
\begin{align}
&G_{\nu,j}(x,t) \nonumber \\
&= \exp \bigg\{ -
\frac{A_{\nu,j}}{2}
\int_0^\infty \d q \frac{\sin^2 qx}{q} \nonumber \\
&\times \left( 1 - \e^{-\im v_j q t} \right)
\left( 1 - \e^{ \im v_j q t} \right)
( 1 + 2 g(v_j q) )
\nonumber \\
& - \frac{B_{\nu,j}}{2} 
\int_0^\infty \d q \frac{\cos^2 qx}{q} \nonumber \\
&\times \left( 1 - \e^{-\im v_j q t} \right)
\left( 1 - \e^{ \im v_j q t} \right)
( 1 + 2 g(v_j q) )
\bigg\}, \\
 &H_{\nu,j}(x,t) \nonumber \\
&= \exp \bigg\{
\frac{A_{\nu,j}}{2}
\int_0^\infty \d q \frac{\sin^2 qx}{q}
\left(
\e^{\im v_j q t} - \e^{- \im v_j q t}
\right)  \nonumber \\
& + \frac{B_{\nu,j}}{2} 
\int_0^\infty \d q \frac{\cos^2 qx}{q}
\left(
\e^{\im v_j q t} - \e^{- \im v_j q t}
\right)
\bigg\},
\end{align}
where
\begin{equation}
 A_{\nu,j} = \frac{(X_{\nu,j})^2}{2 \pi v_j}, \;\;
 B_{\nu,j} = 2 \pi v_j \left[(KX)_{\nu,j}\right]^2,
\end{equation}
and $g(\epsilon) = (\e^{\epsilon/T} -1)^{-1}$ is the Bose distribution
function.
As a result,
\begin{align}
& \langle \left\{ \psi^\dagger_{\nu} (x,0), \psi_{\nu} (x,t)
 \right\}\rangle \nonumber \\
=& \frac{1}{\pi \alpha} 
\exp\left\{
- \sum_{j=1}^N \left[
C_+ I_j(0,t) + C_- I_j(x,t)
\right]
\right\} \nonumber \\
&\times \Bigg[
\exp\left\{
- \sum_{j=1}^N \left[
C_+ J_j(0,t) + C_- J_j(x,t)
\right]
\right\} \nonumber \\
&+ (t \to -t)
\Bigg],
\end{align}
where $C_{\pm} \equiv (B_{\nu,j} \pm A_{\nu,j})/2$ and
\begin{align}
I_j (x,t) =& \int_0^\infty \d q \frac{\cos 2 q x}{q} (1 - \e^{-\im v_j q t})
(1 - \e^{\im v_j q t}) g(v_j q) \nonumber \\
=& \frac{1}{2} \sum_{n=1}^\infty
\Bigg\{
\log \left[ 1 + \frac{(v_j t + 2x)^2}{(n v_j/T)^2}\right]
+ \log \left[ 1 + \frac{(v_j t - 2x)^2}{(n v_j/T)^2}\right] \nonumber \\
& -2 \log \left[ 1 + \frac{(2x)^2}{(n v_j/T)^2}\right]
\Bigg\} \nonumber \\
=& \frac{1}{2}
\Big\{
\log \frac{\sinh \pi T (t + 2x/v_j)}{\pi T (t + 2x/v_j)}
\frac{2 \pi T x/v_j}{\sinh 2 \pi T x/v_j} \nonumber \\
&+ \log \frac{\sinh \pi T (t - 2x/v_j)}{\pi T (t - 2x/v_j)}
\frac{2 \pi T x/v_j}{\sinh 2 \pi T x/v_j}
\Big\},
\\
J_j (x,t) =& \int_0^\infty \d q \frac{\cos 2 q x}{q} (1 - \e^{\im v_j q
 t}) \nonumber \\
=& \frac{1}{2} \left\{
\log \frac{\alpha - \im (v_j t + 2x)}{\alpha - \im 2x}
+ \log \frac{\alpha - \im (v_j t - 2x)}{\alpha + \im 2x}
\right\}.
\end{align}
Since $I_j (\infty,t) = J_j(\infty,t) = 0$,
these results lead to Eqs.~(\ref{eqn:a1})--
(\ref{eqn:aa3dash}).

%
%

\end{document}